\documentclass[11pt,twoside]{article}

\usepackage{asp2006}
\usepackage{graphicx}
\usepackage{lscape}
\usepackage{natbib}

\begin{document}

%
%

\title{Structure and Dynamics of Sunspots}

\author{A. Tritschler}

\affil{National Solar Observatory/Sacramento Peak\footnote{Operated by the %
       Association of Universities for Research in Astronomy, Inc. (AURA), %
       for the National Science Foundation}, P.O.~Box 62, Sunspot, NM-88349, U.S.A.} 

%
%

\begin{abstract}

The physics of Sunspots is a fascinating and demanding field  of
research in solar astronomy. Interaction of magnetic fields and
plasma flows takes place in a tangled magnetic geometry and occurs on
spatial scales that pose a continuous challenge for existing
instrumentation and for the unambiguous interpretation of
spectropolarimetric observations.  Thus, the main properties of
sunspots are well established but its fine structure is not yet fully
understood. 

In this contribution we summarize the current knowledge of the
magnetic and dynamic properties of sunspots at the photospheric level
based on selected observations featuring the highest possible spatial
and spectral resolution. We concentrate on light bridges,  umbral
dots, penumbral filaments and the notorious dark cores in penumbral filaments.  We
report on the morphology of the fine structure elements but mostly
focus on observations of  their line-of-sight velocities and magnetic
field parameters.  We briefly comment on results from recent radiative
MHD simulations and more schematic model ideas that attempt to
rationalize observations of the penumbra. 

\end{abstract}

%
%

\section{Introduction}\label{sec:introduction}

The main properties of a sunspot in terms of flows and magnetic field
have been very well characterized in the last decades. Sunspots
harbour magnetic fields which we now know inhibit at least  partially
convective energy transport leading to the darker appearance of the
umbra which radiates only at 20\,\% of the flux of the quiet sun and the
brighter penumbra which radiates at about 80\,\% of the flux of the quiet
sun. 

Flows in the umbra are mostly negligible except in very young
sunspots \citep{sigwarth+schuessler1998}.
The most vigorous flows are observed in the penumbra
dominated by the Evershed flow which is interpreted as a horizontal
outflow pattern with velocities roughly in the range of 2-6\,km/s.

The magnetic field is almost vertical in the umbra with field
strengths of about 2-3.5\,kG measured by the Zeeman effect and becomes
more horizontal in the penumbra where the field strength drops to
700-1000\,G at the outer penumbral boundary. Forced by the exponential decrease
of density and flux conservation the field lines spread
out with height giving rise to a magnetic canopy. Furthermore
we know that the penumbra carries a significant fraction of the total
magnetic flux which makes it a deep structure as opposed to a shallow
structure \citep{solanki+schmidt1993}.

Besides its large-scale structure that is stable on the time scale of
days to weeks, the penumbra and umbra of sunspots appears to be
organized on very small but dynamic spatial scales ($\sim$0.1\,arcsec).  
We can assume that the magnetic field and its interaction
with plasma flows is the main actor in producing and organizing this
fine structure. So a characterization of this interaction
that is as accurate as possible is important for 
inferring and finally understanding the underlying physics.

%
%

\section{Umbral Fine Structure}\label{sec:umbra}

\subsubsection{Umbral Dots}

The sunspot umbra is filled with tiny brigthenings called umbral dots
(UDs). Depending on their location and brightness inside the umbra
UDs have been either named peripheral UDs (PUDs) or central UDs
(CUDs) \citep{grossmanndoerth+schmidt+schroeter1986}. Recent 
results though suggest a different origin of both species 
\citep{riethmueller+etal2008, sobotka+jurcak2008}.  
On average UDs are about 1000\,K hotter than the coolest parts
of the umbra and only some individual UDs reach or exceed the average
photospheric brightness and temperature. Observed diameters peak at
about 0.23\,arcsec corresponding to about 170\,km \citep{sobotka+hanslmeier2005}. 
Most UDs are known to move towards the more central parts of the umbra or along faint
light bridges with proper motions in the range of several 100\,m/s
\citep{sobotka+puschmann+hamedivafa2008, rimmele2008}. Sometimes UDs are found to
merge or split \citep{kitai+etal2007, sobotka+puschmann+hamedivafa2008}. 
There is also ample evidence that UDs have a substructure in form 
of a dark central lane \citep{rimmele2008}. 

Observed Dopplershifts in UDs vary over a wide range from
unspectacular \citep{schmidt+balthasar1994} to up 
1\,km/s \citep{rimmele2004}. Direct
measurements of a reduction of magnetic field strength in UDs from the
splitting of spectral lines show typically values of 5-10\,\%
\citep{schmidt+balthasar1994} sometimes 20\,\% at the most 
\citep{wiehr+degenhardt1993}. 
Both parameters, Dopplershifts and measured magnetic 
field strength reduction, depend on the range of formation
heights of the spectral line under consideration.

Inversion techniques that allow us to infer information about the
height dependence of the physical  parameters show evidence for
considerably higher upflow velocities (PUDs) and magnetic field
strength reduction  (PUDs, CUDs) in the deep layers (close to the
formation of the continuum) accompanied also by more inclined
magnetic fields that form a (mini-)canopy at least above PUDs
\citep{socas+etal2004, riethmueller+solanki+lagg2008}. The
absence of either enhanced flows and inclined fields in  CUDs is
ascribed to a screening effect caused by their deeper origin. 

\subsubsection{Lightbridges}

Another phenomenon encountered in the umbra are light bridges (LBs):
bright structures that separate umbral cores or
penetrate deeply. Their widths can vary from 
$<$1\,arcsec to several arcsecs across.
Their formation is often observed during either
the decay or formation process of a sunspot. 
LBs appear in very different shapes which might also reflect
their evolutionary stage. They can adopt the form of a narrow 
bright band or of an arrangement of granule-like 
structures giving a segmented and elevated impression, 
which is most obvious when observed out of disk center \citep{lites+etal2004}.
At the highest spatial resolution, however, all LBs have in common 
that they are nerved by a dark central lane associated with a
weak upflow \citep{giordano+etal2008}.
Magnetic fields in LBs are generally weaker \citep[e.g.,][]{katsukawa+etal2007b}
and more inclined with respect to the local vertical \citep[e.g.,][]{leka1997}.
The gathered information raised the suspicion
that LBs constitute a manifestation of 
magnetoconvection in the umbra \citep[e.g.,][]{rimmele2004, 
berger+berdyugina2003, spruit+scharmer2006}. 
This is corroborated by the more recent findings from inversion of 
high-spatial resolution spectropolarimetric observations 
\citep{jurcak+martinezpillet+sobotka2006}. 
These indicate that LBs are a deep phenomenon
formed by field-free (or weak-field) plasma intruding from below
forcing the magnetic field in the umbra to form a canopy
above most parts of the lightbridge.

\subsubsection{Magnetoconvection in the umbra}

A major breakthrough in understanding the umbra and its fine structure has been 
achieved by the three-dimensional radiative MHD simulations
by \cite{schuessler+voegler2006}. The simulations show
the development of nonstationary narrow plumes of rising hot plasma
as a natural result of convection taking 
place in a strong, initially monolithic magnetic field (about 2500\,G). 
The plumes have a significantly reduced magnetic field strength 
in their upper layers (down to a few hundred Gauss) 
and show up in continuum intensity in form of horizontally
elongated structures featuring a central dark lane. 
The central lanes are associated with an upflow while the outer 
end point of the lanes show adjacent downflows.
The dark central lane and its flow signature is interpreted as the direct consequence
of radiative cooling effects which lead to a pile-up of plasma and the formation a
cusp. The flow turns horizontal below the cusp and the locally increased
density and pressure elevates the $\tau = 1$ level to layers 
with a lower temperature causing the appearance of the dark lane.
The elevation of the $\tau = 1$ level can effectively screen
the actual properties of the plume because the formation of spectral lines
takes part in the layers above the cusp. 
It should be noted here that the dark lanes observed 
in LBs likely have a very similar
explanation as the one for UDs \citep{nordlund2006}. 

%
%

\section{Penumbral fine structure}\label{sec:penumbra}

\subsubsection{Penumbral Filaments, Penumbral Grains, Dark-Cored Filaments}

The appearance of the penumbra in intensity is engraved by 
radially aligned alternating bright and dark
structures, the penumbral filaments or fibrils. With good spatial resolution 
the bright filaments appear as a continuous body with a brighter
head of elongated shape, the penumbral grain (PG). In the
inner penumbra PGs and the attached filaments show an inward migration
towards the umbra (0.5 - 1.5\,km/s), while in the outer penumbra an
outward migration is observed, defining a very distinct 
dividing line in the middle penumbra \citep[e.g.,][]{bovelet+wiehr2003, deng+etal2007}.

With the highest spatial resolution the (bright) filament body as well as 
some PGs reveal a conspicuous and striking internal structure. PGs 
appear to be split or crossed \citep{rouppe+etal2004} 
and the filament body is veined by a dark lane \citep{scharmer+etal2002}, respectively. 
The spatial scale of these structures is about 0.12\,arcsec. 
\citet{suetterlin+bellot+schlichenmaier2004} are the first to report
that dark-cored penumbral filaments are mostly observed 
in the inner center-side penumbra. \citet{langhans+etal2007} find
that the visibility of the dark-cored filaments depends on the heliocentric
distance in such a way that it degrades with increasing 
distance from sun center. On the centre-side penumbra, however, 
the dark cores of penumbral filaments are visible for all heliocentric distances. 
Dark-cored penumbral filaments frequently split in the umbra and
form a Y-shape which disappears after a while leaving a shortened
filamentary structure and a bright dot in the umbra.

The dark cores in penumbral filaments are usually best identified in
integrated (1\,nm) G-Band observations at 430.5\,nm taking
advantage of the higher spatial resolution achieved in the blue
wavelength range. However, it appears that they are also more
pronounced in narrowband images taken in the line core
of spectral lines when compared to the local continuum, which 
has been taken as evidence for that these structures are elevated
above the continuum formation height \citep{spruit+scharmer2006, rimmele2008}. 

Another peculiarity inherent to bright filaments is their twisted appearance
caused by intensity fluctuations across the filaments.
The twisting motions are observed only perpendicular to the 
symmetry line, which connects spot center with
sun center. The twist direction is oriented from
limb- to disk-center side. For more details see 
\citet{ichimoto+etal2007a}.

%
%

\section{Flows in the Penumbra}\label{sec:flows}

\subsubsection{Photospheric Evershed Flow}\label{ssec:evershed}

\begin{figure}[t]
  \includegraphics[width=0.49\textwidth]{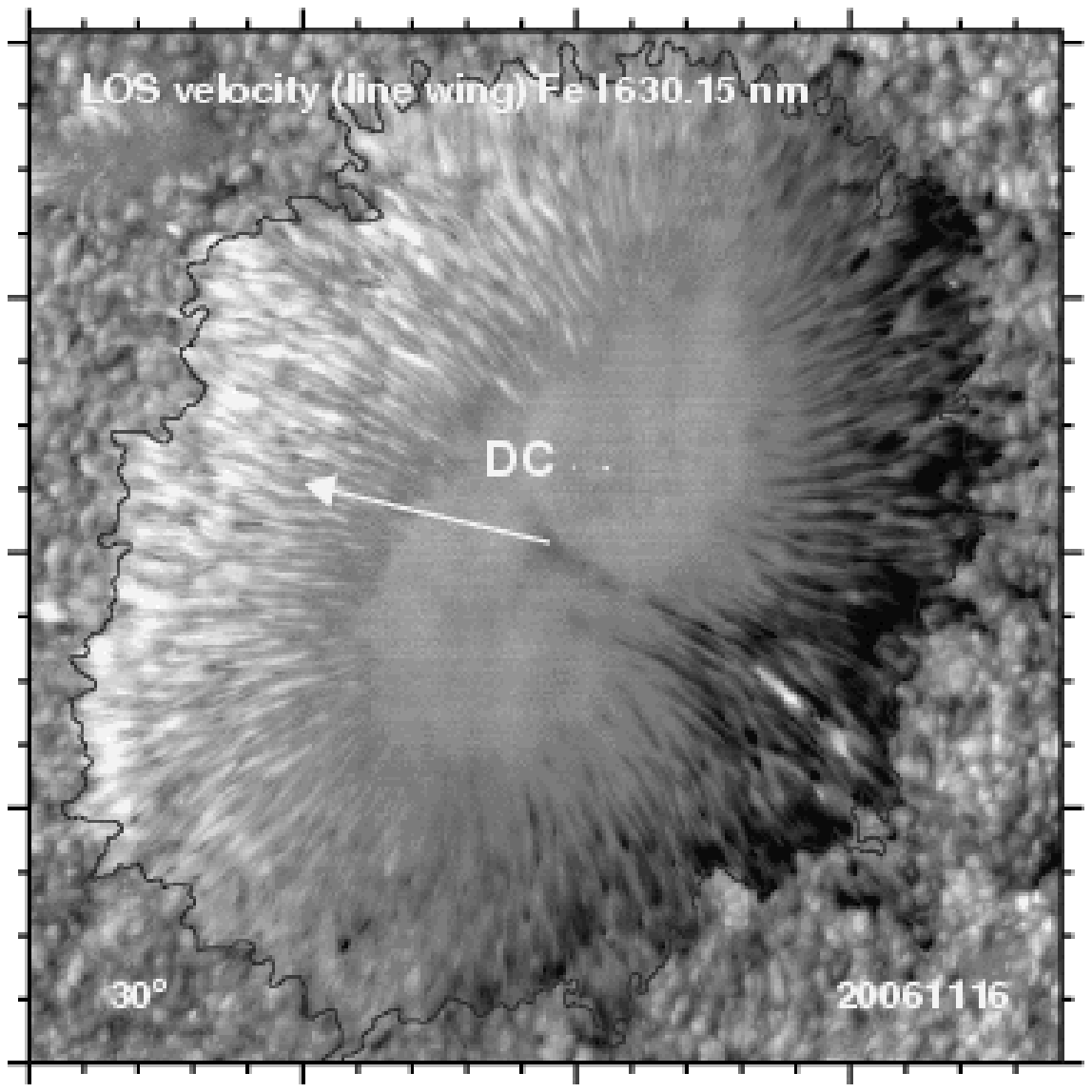}
  \includegraphics[width=0.49\textwidth]{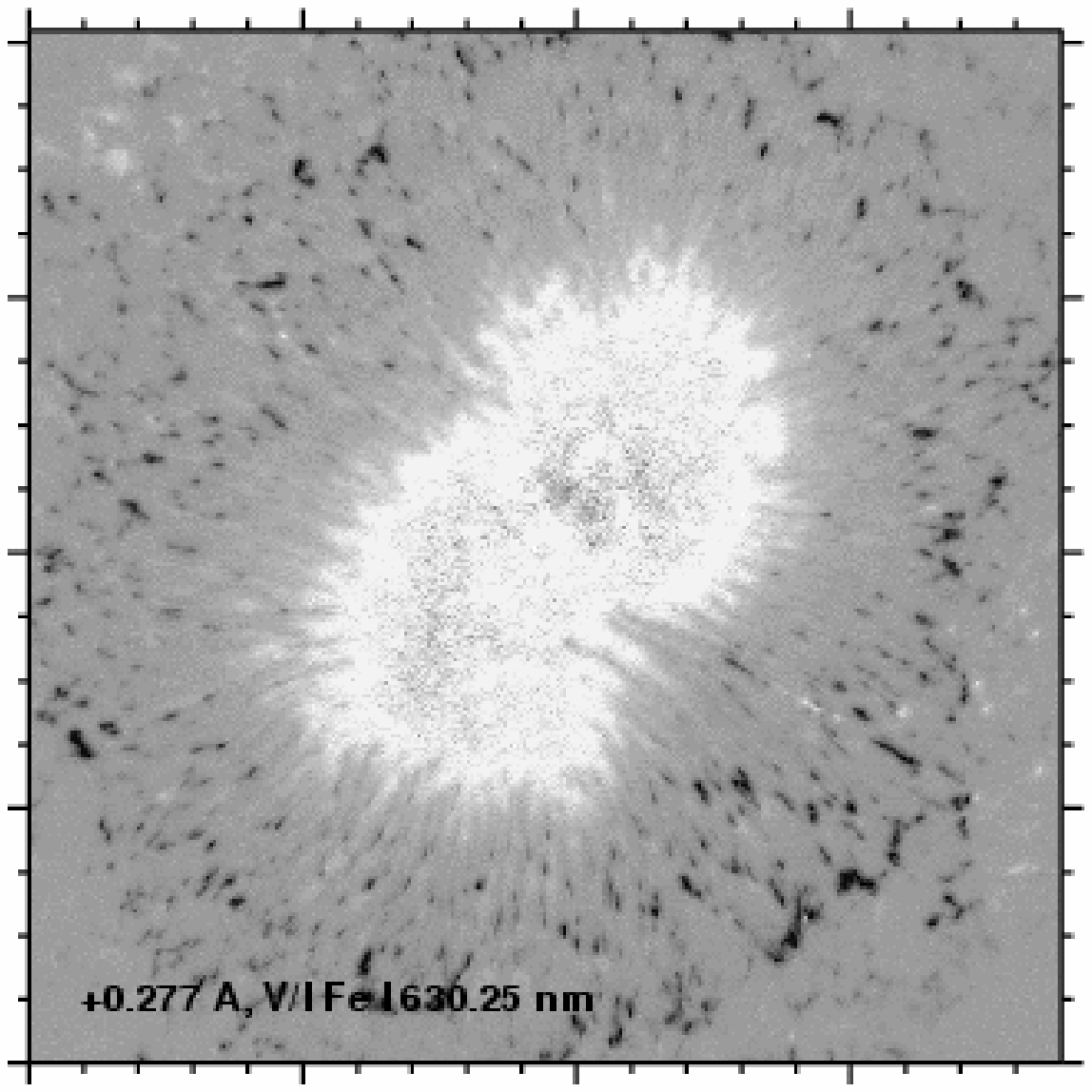}
  \caption{LOS velocity (left) and magnetogram (right)
           of active region NOAA 10923 observed with the spectropolarimeter
           aboard the Japanese satellite HINODE on 
           16 November (out of disk center) and 14 November 
           (disk center), 2006, respectively. 
           The arrow points towards disk center. The LOS velocity
           is determined from the line-wing information of the 630.15\,nm line.
           Bright indicates blueshifts (towards the observer) 
           dark indicates redshifts (away from the observer). 
           Dopplergram is scaled to a velocity range from -2 to 2\,km/s. 
           The magnetogram is taken from the red lobe of the Stokes V signal
           of the 630.25\, nm line. Major tick marks correspond to 20\arcsec.}
  \label{fig:flows}
\end{figure}

The most conspicious and vigorous flow in sunspots is the Evershed
flow \citep{evershed1909}. Inside the penumbra spectral lines appear
blueshifted on the center-side penumbra and redshifted on the
limb-side penumbra. Thus, the Evershed flow is interpreted as a
horizontal and  radially directed outflow.  The Evershed effect does
not only manifest itself in form of systematic Doppler shifts but
also in line asymmetries of spectral lines that depend on the
range of formation heights of the line: weaker (stronger) lines show larger
(smaller) line-core Dopplershifts but less (more) asymmetry. 
For a stronger line the largest line asymmetries and thus
velocities are found in the line wing and close to the continuum
indicating that the Evershed effect is a deep phenomenon 
\citep[e.g.,][]{schlichenmaier+bellot+tritschler2004}. The 
magnitude of the LOS velocity decreases with height 
\citep{boerner+kneer1992, rouppe2002} and 
increases at any given height towards the outer penumbral border where it ceases
rather abruptly \citep[e.g.,][]{hirzberger+kneer2001}.
Furthermore, the location of the maximum shifts to
larger radial distances when higher photospheric layers (e.g., line
core) are probed. To visualize the spatial pattern of the flow field within the penumbra
Figure \ref{fig:flows} (left) displays a Dopplergram using
the line-wing (deeper layers) information of the 630.15\,nm line
as observed with the Solar Optical Telescope \citep[SOT,][]{ichimoto+etal2004}
aboard the Japanese satellite HINODE \citep{kosugi+etal2007}. Noticeable is also the
strong filamentation of the flow field into flow channels, which are
usually better visible and  more pronounced in the center-side
penumbra. 

%
%

\subsubsection{Geometry of the flow field}\label{ssec:geometry}

A full characterization of the flow field, however, is only possible
if the inclination and thus the three dimensional  geometry of the
flow field can be reconstructed from the observed spectral
information. The geometry question is of crucial importance  to the
more general question that is raised when the Evershed effect  is
interpreted as a horizontal mass outflow: where are the sources  and
the sinks of the flow? The apparent asymmetry between the spatial
location  of largest blueshifts on the center-side and redshifts  on
the limb-side penumbra already is an indication 
that the flow must be inclined. From methods that
determine the average flow geometry it has been
inferred that in deep layers  the average inclination of the flow
vector changes with radial distance: it is slightly inclined upwards
in the inner penumbra, than becomes horizontal and further outwards
returns back to the solar surface \citep[e.g.][]{tritschler+etal2004}. 
These results  are in agreement with findings that employ a more 
sophisticated approach like an inversion \citep{bellot+etal2003}. 
In higher layers some  part of the flow ($\sim$10-20\,\%) continues 
upwards to the  magnetic canopy 
\citep{solanki+montavon+livingston1994, rezaei+etal2006}. 

The upflow and downflow component of the flow field, as suggested
by the average geometry, are also observed 
\citep[e.g.,][]{rimmele1995b, westendorp+etal1997, schlichenmaier+schmidt1999}. 
Although this vertical component of the Evershed
flow is best and unambiguously identified only in observations at or
very close to disk center, evidence also comes from observations 
of the limb-side penumbra at larger viewing angles. The latter 
clearly show that the body of the filament
flow channels on the limb-side penumbra is mostly associated with a
redshift (the Evershed outflow)  but the head of the filament (the
penumbral grain) shows a very distinct and spatially confined
blueshift interpreted as an upflow \citep{rimmele+marino2006}. 
The turnover between upflow and outflow takes place on 
spatial scales much less than 1\,arcsec.

The existence of up and downflows is not limited to the inner and
outer penumbra as is demonstrated by magnetograms taken in the
extreme red (see Figure \ref{fig:flows}, right) and 
blue line wing of the Stokes-V signal: tiny
elongated structures associated with a small upflow are observed
throughout the penumbra and extended downflow patches are preferrably
found at the outer boundary but not exclusively 
\citep{ichimoto+etal2007a}. Many or most of the
downflow patches have opposite polarity from the sunspot. The
magnetogram signal is co-spatial with LOS Dopplergram signal which
reassures that real velocity  signals are detected but more
importantly  illustrates clearly that the flows occur in a
magnetized medium.  However, it is difficult to identify pairs
of upflows and downflows which leaves the question of 
mass balance still an open issue. 

\subsubsection{Velocity-intensity correlation}

The obvious filamentation of the flow immediatly raises the suspicion
that the Evershed flow channels must be coupled to the
penumbral fine structure as it is observed also in intensity.  A
positive correlation between intensity and flows (like it is the case
for granulation) has been used as an argument for a convective origin
of the filamentation altered only by the presence of magnetic fields.
In principle it seems plausible that any outcome of such a correlation
analysis depends critically on the spatial resolution (besides other
factors like e.g. spectral resolution, orientation, radial distance
from umbra) achieved in the spectroscopic or spectropolarimetric
observations and whether the contribution of the dark cores of penumbral filaments
is clearly resolved or not.

In hindsight, the lack of spatial resolution can explain most of the
variety of findings of the many observations that attempted to address
this question. Observations hint towards either a correlation of
flows with dark intensity features 
\citep{westendorp+etal2001a, rouppe2002}, no significant correlation 
at all \citep[e.g.][]{hirzberger+kneer2001} or a
correlation of stronger flows with bright (dark) structures in the
inner (middle and outer) penumbra \citep{jurcak+etal2007,jurcak+bellot2008,
ichimoto+etal2007a}. In some cases flow channels connect bright and
dark features \citep{schlichenmaier+bellot+tritschler2005}.

The best available data shows that the Evershed effect seems to be
concentrated in the dark cores of penumbral filaments 
\citep{bellot+langhans+schlichenmaier2005, 
rimmele+marino2006, langhans+etal2007, bellot+etal2007}. 
\citet{bellot+langhans+schlichenmaier2005} 
demonstrate spectroscopically via bisector shapes that dark cores 
(center side) are associated with strong line-wing blueshifts  while the
corresponding lateral brightenings show much less LOS velocity
signal. It is fair to note that sometimes dark cores exhibit just the
same LOS velocity as their flanking brightenings. In a so far unique
observation by \cite{rimmele2008}, the LOS Dopplergram of a dark-cored
filament channel reveals blueshifts (upflows) associated with the dark
core and redshifts (downflows) in the brightenings when deeper layers
(line wing) are probed. In higher layers (line core) the same filament
shows a blueshift confined to the bright head of the filament and a
redshift along the whole filament body without any signature of the
presence of the dark core. Whether this observation represents just a
one-time peculiarity or describes a common behaviour still needs to
be verified.

%
%

%
\begin{figure}[t]
  \includegraphics[width=0.49\textwidth]{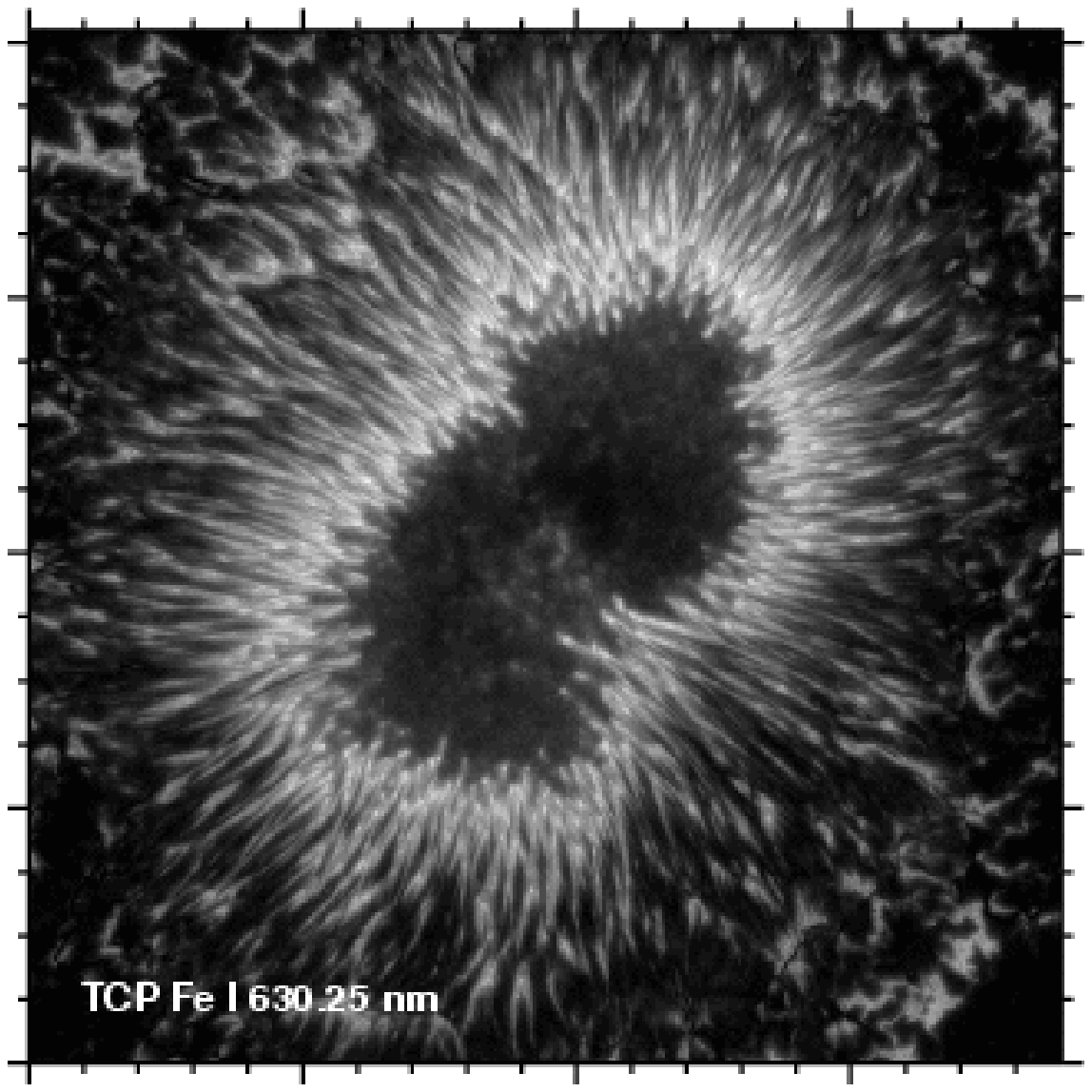}
  \includegraphics[width=0.49\textwidth]{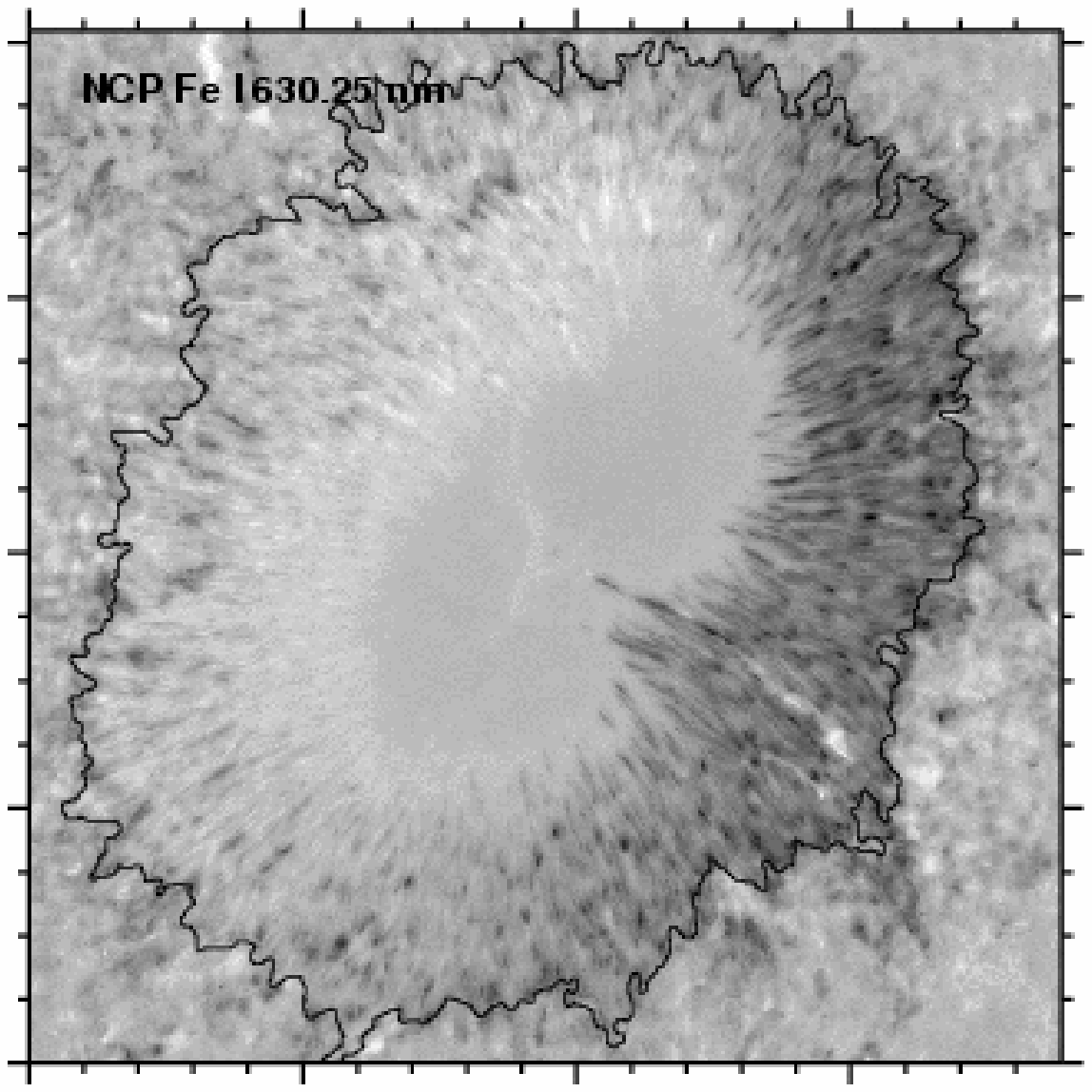}
  \caption{Total circular polarization (left) and NCP (right) 
           of active region NOAA 10923 observed in the 
           Fe~I 630.25\,nm line with the spectropolarimeter
           onboard the Japanese satellite HINODE on November 
           14 (disk center) and November 16 
           (out of disk center), 2006, respectively. 
           Major tick marks correspond to 20\arcsec.}
  \label{fig:magnetic}
\end{figure}
%

\section{Magnetic Fields in the Penumbra}

\subsubsection{Fluted Penumbra}

Sunspots do not only show a filamentary structure in intensity and the
flow field but also in polarized light as visualized 
in Figure \ref{fig:magnetic}(left). 
The azimuthal inhomogeneity of the penumbral magnetic field on small
scales is well established going back to the early findings of 
\cite{beckers+schroeter1969b}. The penumbra appears to be "fluted" on small
spatial scales \citep{title+etal1993, langhans+etal2005}.
A different signal in circular polarization can either indicate different
magnetic field strength or a different inclination
and it needs either an inversion of spectropolarimetric 
data or some other technique to disentangle these two effects. 
It turns out that two components can be separated: one which 
harbours more vertical and stronger fields, 
and one showing weaker and more inclined fields, which have been named
spines and intra-spines, respectively \citep{lites+etal1993, 
borrero+solanki2008, borrero+lites+solanki2008}. For a very
thorough characterization of these two magnetic components 
see also \citet{bellot+balthasar+collados2004}.
In general, the conclusion is that the azimuthal fluctuation
of polarization is caused by a combined effect of varying field
inclination and field strength. Furthermore, the dark cores in penumbral filaments tend to
have more inclined and weaker fields and also are the location where 
the Evershed flow is concentrated \citep{langhans+etal2007, bellot+etal2007}. 

\subsubsection{Stokes Asymmetries, NCP}

An alternative and more direct way to extract 
information about how the magnetic field is related to 
flows, is to examine the asymmetry of the
Stokes parameters with wavelength \citep[e.g.,][]{sanchezalmeida+lites1992}. 
In particular the wavelength integrated Stokes V profile, 
the net-circular polarization (NCP),
provides a simple and instructive way of characterization.
The NCP, although a condensed quantity, is of
diagnostic importance because apart from the Evershed effect it is a very
critical observation that any model of the penumbra must reproduce.
The spatial behaviour of the NCP is symmetric in the inner and middle penumbra
w.r.t. the line that connects the disk center with the spot
center (see Figure \ref{fig:magnetic}, right). 
The same property applies to the LOS velocity which
already indicates an intricate coupling between penumbral flows and the sign
of the NCP. This is corroborated by the fine structure of
the NCP, which reveals filamentary organized inhomogeneities throughout
the penumbra with very localized enhancements of elongated shape on
very small spatial scales. Furthermore, there appears to be a
difference in the symmetry properties of the NCP observed between the
limb- and center-side penumbra, namely a sign reversal of the
NCP in the outer center-side penumbra \citep{tritschler+etal2007}.
In anticipation of a coupling between flows and the NCP
ground-based observations have had difficulties to establish 
suchlike conclusively. However, seeing-free spectropolarimetry 
suggests a correlation between the two quantities. 
The positive NCP is co-spatial with flow channels on the
center- and limb-side penumbra, while a negative NCP
is co-spatial with locations in between the 
flow channels \citep{ichimoto+etal2008}.

%
%

\section{Schematic Models of the Penumbra}

\subsubsection{Uncombed Structure}

In an attempt to master some of the observations, particularly 
the asymmetries of the Stokes parameters and the azimuthal 
variation of the polarization signal, \citet{solanki+montavon1993} 
proposed the concept of the uncombed penumbra.
It is the simplified idea of more or less horizontal flux tubes 
that carry the Evershed flow and which are embedded in a more 
vertical background magnetic field.
There exist different realizations of the uncombed penumbra but in all
these models sharp gradients in velocity and magnetic field parameters
are encountered at the interface between the flux tube and the background
that can contribute to the generation of the observed NCP.
Velocity gradients are a necessary condition though to produce a NCP.
The early two-dimensional models 
\citep{solanki+montavon1993, martinezpillet2000} 
do reproduce observed properties 
of the NCP but are limited to a few geometric cases. More importantly, 
they cannot provide any information about the spatial distribution of
the NCP within the penumbra. In a more sophisticated approach 
this downside was overcome by a three-dimensional 
model incorporating a snaphsot of the moving tube model embedded in a
static background atmosphere \citep{mueller+etal2002, mueller+etal2006}. 
In particular, this approach allowed to identify
the discontinuity in the apparent azimuth of the magnetic field 
(between flux tube and background as measured in the observers 
coordinate system) and anomalous dispersion as the main cause of the characteristic spatial 
pattern of the NCP when observed in the infrared at 1564.8 nm. 

\subsubsection{Field-Free Gaps in the Penumbra}

In order to explain the observed brightness of
the penumbra and inspired by the striking morphological 
similarities between UDs,
LBs and penumbral filaments (dark central lanes), 
\citet{spruit+scharmer2006} and \citet{scharmer+spruit2006} proposed 
and investigated radially aligned field-free plumes where convection 
can take place thus promoting an all-embracing picture of the fine structure
in sunspots. The gappy penumbra has since then challenged the
concept of penumbral flux tubes. The magnetohydrostatic model predicts
a field topology above the field-free gaps that changes from 
cusp-shaped in the inner to canopy-shaped in the outer penumbra.
Hence, the field geometry above the gaps together with the 
convective flows inside the field-free gaps provides gradients and discontinuities
that in principle could lead to a NCP, but unfortunately this has not
been demonstrated yet. The Evershed flow is interpreted as the
horizontal flow component of convective motions in the field-free gaps.
For the most recent discussion of this approach see \citet{scharmer2008}.

Obviously, a crucial question that must be answered 
by any model of the penumbra is how the observed penumbral heat flux 
($\sim$75\,\% of the quiet sun) can be accounted for. 
Convective motions in field-free gaps solve this problem in 
a natural way while on the other hand, the hypothetical interchange convection 
\citep[e.g.,][]{jahn+schmidt1994} alone cannot acount 
for the brightness of the penumbra \citep{schlichenmaier+solanki2003}.
However, in an extension of the work performed 
by \citet{schlichenmaier+bruls+schuessler1999}, 
\cite{ruizcobo+bellot2008} demonstrate that thick flux tubes
carrying a (hot) Evershed flow can explain the surplus brightness of
the penumbra \citep[see also][]{schlichenmaier+solanki2003}.

\subsubsection{Observational evidence}

In either model, the uncombed or gappy penumbra, the magnetic 
field has to close or wrap around either the embedded flux tube or 
the field-free gap which would show up in the form
of an opposite sign of the azimuth angle on both sides above the
penumbral filaments. Such field wrapping has been recently
inferred from inversions of seeing-free spectropolarimetry
further corroborating the strong spinal (stronger fields) and intra-spinal 
(weaker fields) structure of the magnetic field 
\citep{borrero+lites+solanki2008}. 

A type of roll-over convection in penumbral filaments is mediated
by the aforementioned apparent twist that is seen 
in continuum intensity space-time plots taken across the filaments
\citep{ichimoto+etal2007b}. The character of that twist 
strongly suggests that it is the result of a viewing angle effect 
instead of being caused by an actual helical motion of the filaments 
\citep[see also][]{zakharov+etal2008}. For a different 
interpretation see \cite{ryutova+berger+title2008}.

In deeper layers \cite{rimmele2008} directly observed redshifts associated with
the bright flanks of a dark penumbral core showing blueshifts. 
In higher layers the same filament shows blueshifts in the PG
and reshifts along the filament body. Notwithstanding the so far unique
but also ambiguous nature of this observation, it must be viewed as
evidence for convection in penumbral filaments. 

%
%

\section{Simulations of Penumbral Fine Structure}

Sunspots pose a very challenging 
regime for modelers studying magneto-convection.
The situation for the penumbral fine structure 
is particularly complicated by the presence of oblique 
fields and the larger spatial scales (horizontal and vertical) 
that must be taken into account. 
All the more impressive appear the most recent attempts of
\citet{heinemann+etal2007} and \citet{rempel+schuessler+knoelker2009} 
that try to tackle the difficult regime
encountered in the penumbra. Both numerical simulations 
encompass three-dimensional MHD including
grey radiative transfer. 

Apart from morphological differences 
both simulations in principle show similar results like
the development of more or less evolved filamentary stuctures
established by gaps with reduced field strength. The horizontal
component of overturning or roll-type convection  in these gaps
provides an explanation for the Evershed flow. The flow speeds,
however, are much smaller than the actual observed ones. The dark-cored
structures that are associated with weaker and more horizontal fields
harbour an outflow and migrate inwards. At the outer boundary of
of this rudimentary penumbra fragments of flux are carried away by a
large scale moat flow that develops as well during the
simulation. It is yet very striking, that although individual filaments do
develop, there is no evolved penumbra, meaning the penumbra 
is not filled  with filaments. This is currently not understood.

%
%

\section{Concluding Remarks} 

We finally see exciting simulations of radiative MHD
coming to a realistic state including also the penumbra. 
A very important step will be to demonstrate that these
simulations can reproduce the observations not only from a
morphological standpoint but also from a spectroscopic perspective. 
There is a wealth of spectropolarimetric information that needs
to be explained. Therefore, it is the next logical step 
to perform spectral synthesis based on these simulations 
to allow for a direct comparison with the observations 
in the important wavelength bands of the visible 
and the infrared. 

In the mean-time we have to rely further on high-resolution observations
(spectrally, spatially and temporally) provided by improved
instrumentation, image reconstruction techniques and interpretational
tools like inversion codes which are necessary to derive the magnetic
field geometry while nevertheless keeping in mind the limitations of their
diagnostic capabilities. 

In a future attempt the photospheric structure must be tied 
to the chromospheric structure of sunspots and important questions like 
(a) what is the cause of the inverse Evershed effect, (b) what
is the magnetic field structure in the chromosphere and (c) how does it
relate to the uncombed fields observed in the photosphere? must be adressed. 

With the launch of HINODE a stable window to the Sun
opened, providing a spatial resolution close 
to the diffraction limit of its telescope.
Hence, the joint effort of ground-based and space-borne observations
should lead to substantial progress in characterization 
of the various small-scale inhomogeneities observed in sunspots. 

%
%


\begin{thebibliography}{}

\bibitem[{{Beckers} \& {Schr{\"o}ter}(1969)}]{beckers+schroeter1969b}
{Beckers}, J.~M., \& {Schr{\"o}ter}, E.~H. 1969, \solphys, 10, 384

\bibitem[{{Bellot Rubio} {et~al.}(2004){Bellot Rubio}, {Balthasar}, \&
  {Collados}}]{bellot+balthasar+collados2004}
{Bellot Rubio}, L.~R., {Balthasar}, H., \& {Collados}, M. 2004, \aap, 427, 319

\bibitem[{{Bellot Rubio} {et~al.}(2003){Bellot Rubio}, {Balthasar}, {Collados},
  \& {Schlichenmaier}}]{bellot+etal2003}
{Bellot Rubio}, L.~R., {Balthasar}, H., {Collados}, M., \& {Schlichenmaier}, R.
  2003, \aap, 403, L47

\bibitem[{{Bellot Rubio} {et~al.}(2005){Bellot Rubio}, {Langhans}, \&
  {Schlichenmaier}}]{bellot+langhans+schlichenmaier2005}
{Bellot Rubio}, L.~R., {Langhans}, K., \& {Schlichenmaier}, R. 2005, \aap, 443,
  L7

\bibitem[{{Bellot Rubio} {et~al.}(2007){Bellot Rubio}, {Tsuneta}, {Ichimoto},
  {Katsukawa}, {Lites}, {Nagata}, {Shimizu}, {Shine}, {Suematsu}, {Tarbell},
  {Title}, \& {del Toro Iniesta}}]{bellot+etal2007}
{Bellot Rubio}, L.~R., {Tsuneta}, S., {Ichimoto}, K., {Katsukawa}, Y., {Lites},
  B.~W., {Nagata}, S., et al. 2007, \apjl,
  668, L91

\bibitem[{{Berger} \& {Berdyugina}(2003)}]{berger+berdyugina2003}
{Berger}, T.~E., \& {Berdyugina}, S.~V. 2003, \apjl, 589, L117

\bibitem[{{Boerner} \& {Kneer}(1992)}]{boerner+kneer1992}
{Boerner}, P., \& {Kneer}, F. 1992, \aap, 259, 307

\bibitem[{{Borrero} {et~al.}(2008){Borrero}, {Lites}, \&
  {Solanki}}]{borrero+lites+solanki2008}
{Borrero}, J.~M., {Lites}, B.~W., \& {Solanki}, S.~K. 2008, \aap, 481, L13

\bibitem[{{Borrero} \& {Solanki}(2008)}]{borrero+solanki2008}
{Borrero}, J.~M., \& {Solanki}, S.~K. 2008, \apj, 687, 668

\bibitem[{{Bovelet} \& {Wiehr}(2003)}]{bovelet+wiehr2003}
{Bovelet}, B., \& {Wiehr}, E. 2003, \aap, 412, 249

\bibitem[{{Deng} {et~al.}(2007){Deng}, {Choudhary}, {Tritschler}, {Denker},
  {Liu}, \& {Wang}}]{deng+etal2007}
{Deng}, N., {Choudhary}, D.~P., {Tritschler}, A., {Denker}, C., {Liu}, C., \&
  {Wang}, H. 2007, \apj, 671, 1013

\bibitem[{{Evershed}(1909)}]{evershed1909}
{Evershed}, J. 1909, \mnras, 69, 454

\bibitem[{{Giordano} {et~al.}(2008){Giordano}, {Berrilli}, {Del Moro}, \&
  {Penza}}]{giordano+etal2008}
{Giordano}, S., {Berrilli}, F., {Del Moro}, D., \& {Penza}, V. 2008, \aap, 489,
  747

\bibitem[{{Grossmann-Doerth} {et~al.}(1986){Grossmann-Doerth}, {Schmidt}, \&
  {Schroeter}}]{grossmanndoerth+schmidt+schroeter1986}
{Grossmann-Doerth}, U., {Schmidt}, W., \& {Schroeter}, E.~H. 1986, \aap, 156,
  347

\bibitem[{{Heinemann} {et~al.}(2007){Heinemann}, {Nordlund}, {Scharmer}, \&
  {Spruit}}]{heinemann+etal2007}
{Heinemann}, T., {Nordlund}, {\AA}., {Scharmer}, G.~B., \& {Spruit}, H.~C.
  2007, \apj, 669, 1390

\bibitem[{{Hirzberger} \& {Kneer}(2001)}]{hirzberger+kneer2001}
{Hirzberger}, J., \& {Kneer}, F. 2001, \aap, 378, 1078

\bibitem[{{Ichimoto} {et~al.}(2007{\natexlab{a}}){Ichimoto}, {Shine}, {Lites},
  {Kubo}, {Shimizu}, {Suematsu}, {Tsuneta}, {Katsukawa}, {Tarbell}, {Title},
  {Nagata}, {Yokoyama}, \& {Shimojo}}]{ichimoto+etal2007a}
{Ichimoto}, K., {Shine}, R.~A., {Lites}, B., {Kubo}, M., {Shimizu}, T.,
  {Suematsu}, Y., et al. 2007{\natexlab{a}},
  \pasj, 59, 593

\bibitem[{{Ichimoto} {et~al.}(2007{\natexlab{b}}){Ichimoto}, {Suematsu},
  {Tsuneta}, {Katsukawa}, {Shimizu}, {Shine}, {Tarbell}, {Title}, {Lites},
  {Kubo}, \& {Nagata}}]{ichimoto+etal2007b}
{Ichimoto}, K., {Suematsu}, Y., {Tsuneta}, S., {Katsukawa}, Y., {Shimizu}, T.,
  {Shine}, R.~A., et al. 2007{\natexlab{b}}, Science, 318, 1597

\bibitem[{{Ichimoto} {et~al.}(2008){Ichimoto}, {Tsuneta}, {Suematsu},
  {Katsukawa}, {Shimizu}, {Lites}, {Kubo}, {Tarbell}, {Shine}, {Title}, \&
  {Nagata}}]{ichimoto+etal2008}
{Ichimoto}, K., {Tsuneta}, S., {Suematsu}, Y., {Katsukawa}, Y., {Shimizu}, T.,
  {Lites}, B.~W., {Kubo}, M., {Tarbell}, T.~D., {Shine}, R.~A., {Title}, A.~M.,
  \& {Nagata}, S. 2008, \aap, 481, L9

\bibitem[{{Ichimoto} {et~al.}(2004){Ichimoto}, {Tsuneta}, {Suematsu},
  {Shimizu}, {Otsubo}, {Kato}, {Noguchi}, {Nakagiri}, {Tamura}, {Katsukawa},
  {Kubo}, {Sakamoto}, {Hara}, {Minesugi}, {Ohnishi}, {Saito}, {Kawaguchi},
  {Matsushita}, {Nakaoji}, {Nagae}, {Sakamoto}, {Hasuyama}, {Mikami},
  {Miyawaki}, {Sakurai}, {Kaido}, {Horiuchi}, {Shimada}, {Inoue}, {Mitsutake},
  {Yoshida}, {Takahara}, {Takeyama}, {Suzuki}, \& {Abe}}]{ichimoto+etal2004}
{Ichimoto}, K., {Tsuneta}, S., {Suematsu}, Y., {Shimizu}, T., {Otsubo}, M.,
  {Kato}, Y., et al. 2004, Proceedings of the SPIE, 
   Volume 5487, ed. J.~C. {Mather}, 1142

\bibitem[{{Jahn} \& {Schmidt}(1994)}]{jahn+schmidt1994}
{Jahn}, K., \& {Schmidt}, H.~U. 1994, \aap, 290, 295

\bibitem[{{Jurc{\'a}k} {et~al.}(2007){Jurc{\'a}k}, {Bellot Rubio}, {Ichimoto},
  {Katsukawa}, {Lites}, {Nagata}, {Shimizu}, {Suematsu}, {Tarbell}, {Title}, \&
  {Tsuneta}}]{jurcak+etal2007}
{Jurc{\'a}k}, J., {Bellot Rubio}, L., {Ichimoto}, K., {Katsukawa}, Y., {Lites},
  B., {Nagata}, S., et al. 2007, \pasj, 59, 601

\bibitem[{{Jur{\v c}{\'a}k} \& {Bellot Rubio}(2008)}]{jurcak+bellot2008}
{Jur{\v c}{\'a}k}, J., \& {Bellot Rubio}, L.~R. 2008, \aap, 481, L17

\bibitem[{{Jur{\v c}{\'a}k} {et~al.}(2006){Jur{\v c}{\'a}k}, {Mart{\'{\i}}nez
  Pillet}, \& {Sobotka}}]{jurcak+martinezpillet+sobotka2006}
{Jur{\v c}{\'a}k}, J., {Mart{\'{\i}}nez Pillet}, V., \& {Sobotka}, M. 2006,
  \aap, 453, 1079

\bibitem[{{Katsukawa} {et~al.}(2007){Katsukawa}, {Yokoyama}, {Berger},
  {Ichimoto}, {Kubo}, {Lites}, {Nagata}, {Shimizu}, {Shine}, {Suematsu},
  {Tarbell}, {Title}, \& {Tsuneta}}]{katsukawa+etal2007b}
{Katsukawa}, Y., {Yokoyama}, T., {Berger}, T.~E., {Ichimoto}, K., {Kubo}, M.,
  {Lites}, B., et al. 2007, \pasj, 59, 577

\bibitem[{{Kitai} {et~al.}(2007){Kitai}, {Watanabe}, {Nakamura}, {Otsuji},
  {Matsumoto}, {Ueno}, {Nagata}, {Shibata}, {Muller}, {Ichimoto}, {Tsuneta},
  {Suematsu}, {Katsukawa}, {Shimizu}, {Tarbell}, {Shine}, {Title}, \&
  {Lites}}]{kitai+etal2007}
{Kitai}, R., {Watanabe}, H., {Nakamura}, T., {Otsuji}, K., {Matsumoto}, T.,
  {Ueno}, S., et al. 2007, \pasj, 59, 585

\bibitem[{{Kosugi} {et~al.}(2007){Kosugi}, {Matsuzaki}, {Sakao}, {Shimizu},
  {Sone}, {Tachikawa}, {Hashimoto}, {Minesugi}, {Ohnishi}, {Yamada}, {Tsuneta},
  {Hara}, {Ichimoto}, {Suematsu}, {Shimojo}, {Watanabe}, {Shimada}, {Davis},
  {Hill}, {Owens}, {Title}, {Culhane}, {Harra}, {Doschek}, \&
  {Golub}}]{kosugi+etal2007}
{Kosugi}, T., {Matsuzaki}, K., {Sakao}, T., {Shimizu}, T., {Sone}, Y.,
  {Tachikawa}, S., et al. 2007, \solphys, 243, 3

\bibitem[{{Langhans} {et~al.}(2007){Langhans}, {Scharmer}, {Kiselman}, \&
  {L{\"o}fdahl}}]{langhans+etal2007}
{Langhans}, K., {Scharmer}, G.~B., {Kiselman}, D., \& {L{\"o}fdahl}, M.~G.
  2007, \aap, 464, 763

\bibitem[{{Langhans} {et~al.}(2005){Langhans}, {Scharmer}, {Kiselman},
  {L{\"o}fdahl}, \& {Berger}}]{langhans+etal2005}
{Langhans}, K., {Scharmer}, G.~B., {Kiselman}, D., {L{\"o}fdahl}, M.~G., \&
  {Berger}, T.~E. 2005, \aap, 436, 1087

\bibitem[{{Leka}(1997)}]{leka1997}
{Leka}, K.~D. 1997, \apj, 484, 900

\bibitem[{{Lites} {et~al.}(1993){Lites}, {Elmore}, {Seagraves}, \&
  {Skumanich}}]{lites+etal1993}
{Lites}, B.~W., {Elmore}, D.~F., {Seagraves}, P., \& {Skumanich}, A.~P. 1993,
  \apj, 418, 928

\bibitem[{{Lites} {et~al.}(2004){Lites}, {Scharmer}, {Berger}, \&
  {Title}}]{lites+etal2004}
{Lites}, B.~W., {Scharmer}, G.~B., {Berger}, T.~E., \& {Title}, A.~M. 2004,
  \solphys, 221, 65

\bibitem[{{Mart{\'\i}nez Pillet}(2000)}]{martinezpillet2000}
{Mart{\'\i}nez Pillet}, V. 2000, A\&A, 361, 734

\bibitem[{{M{\"u}ller} {et~al.}(2006){M{\"u}ller}, {Schlichenmaier}, {Fritz},
  \& {Beck}}]{mueller+etal2006}
{M{\"u}ller}, D.~A.~N., {Schlichenmaier}, R., {Fritz}, G., \& {Beck}, C. 2006,
  \aap, 460, 925

\bibitem[{{M{\"u}ller} {et~al.}(2002){M{\"u}ller}, {Schlichenmaier}, {Steiner},
  \& {Stix}}]{mueller+etal2002}
{M{\"u}ller}, D.~A.~N., {Schlichenmaier}, R., {Steiner}, O., \& {Stix}, M.
  2002, \aap, 393, 305

\bibitem[{{Nordlund}(2006)}]{nordlund2006}
{Nordlund}, {\AA}. 2006, in ASP Conference
  Series, Vol. 354, Solar MHD Theory and Observations: A High Spatial
  Resolution Perspective, ed. J.~{Leibacher}, R.~F. {Stein}, \&
  H.~{Uitenbroek}, 353--+

\bibitem[{{Rempel} {et~al.}(2009){Rempel}, {Sch{\"u}ssler}, \&
  {Kn{\"o}lker}}]{rempel+schuessler+knoelker2009}
{Rempel}, M., {Sch{\"u}ssler}, M., \& {Kn{\"o}lker}, M. 2009, \apj, 691, 640

\bibitem[{{Rezaei} {et~al.}(2006){Rezaei}, {Schlichenmaier}, {Beck}, \& {Bellot
  Rubio}}]{rezaei+etal2006}
{Rezaei}, R., {Schlichenmaier}, R., {Beck}, C., \& {Bellot Rubio}, L.~R. 2006,
  \aap, 454, 975

\bibitem[{{Riethm{\"u}ller} {et~al.}(2008{\natexlab{a}}){Riethm{\"u}ller},
  {Solanki}, \& {Lagg}}]{riethmueller+solanki+lagg2008}
{Riethm{\"u}ller}, T.~L., {Solanki}, S.~K., \& {Lagg}, A. 2008{\natexlab{a}},
  \apjl, 678, L157

\bibitem[{{Riethm{\"u}ller} {et~al.}(2008{\natexlab{b}}){Riethm{\"u}ller},
  {Solanki}, {Zakharov}, \& {Gandorfer}}]{riethmueller+etal2008}
{Riethm{\"u}ller}, T.~L., {Solanki}, S.~K., {Zakharov}, V., \& {Gandorfer}, A.
  2008{\natexlab{b}}, \aap, 492, 233

\bibitem[{{Rimmele}(2008)}]{rimmele2008}
{Rimmele}, T. 2008, \apj, 672, 684

\bibitem[{{Rimmele} \& {Marino}(2006)}]{rimmele+marino2006}
{Rimmele}, T., \& {Marino}, J. 2006, \apj, 646, 593

\bibitem[{{Rimmele}(1995)}]{rimmele1995b}
{Rimmele}, T.~R. 1995, \apj, 445, 511

\bibitem[{{Rimmele}(2004)}]{rimmele2004}
---. 2004, \apj, 604, 906

\bibitem[{{Rouppe van der Voort}(2002)}]{rouppe2002}
{Rouppe van der Voort}, L.~H.~M. 2002, \aap, 389, 1020

\bibitem[{{Rouppe van der Voort} {et~al.}(2004){Rouppe van der Voort}, {L{\"
  o}fdahl}, {Kiselman}, \& {Scharmer}}]{rouppe+etal2004}
{Rouppe van der Voort}, L.~H.~M., {L{\" o}fdahl}, M.~G., {Kiselman}, D., \&
  {Scharmer}, G.~B. 2004, \aap, 414, 717

\bibitem[{{Ruiz Cobo} \& {Bellot Rubio}(2008)}]{ruizcobo+bellot2008}
{Ruiz Cobo}, B., \& {Bellot Rubio}, L.~R. 2008, \aap, 488, 749

\bibitem[{{Ryutova} {et~al.}(2008){Ryutova}, {Berger}, \&
  {Title}}]{ryutova+berger+title2008}
{Ryutova}, M., {Berger}, T., \& {Title}, A. 2008, \apj, 676, 1356

\bibitem[{{S\'anchez Almeida} \& {Lites}(1992)}]{sanchezalmeida+lites1992}
{S\'anchez Almeida}, J., \& {Lites}, B.~W. 1992, ApJ, 398, 359

\bibitem[{{Scharmer}(2008)}]{scharmer2008}
{Scharmer}, G.~B. 2008, Physica Scripta Volume T, 133, 014015

\bibitem[{Scharmer {et~al.}(2002)Scharmer, Gudiksen, Kiselman, L{\"o}fdahl, \&
  Rouppe van~der Voort}]{scharmer+etal2002}
Scharmer, G.~B., Gudiksen, B.~V., Kiselman, D., L{\"o}fdahl, M.~G., \& Rouppe
  van~der Voort, L.~H.~M. 2002, \nat, 420, 151

\bibitem[{{Scharmer} \& {Spruit}(2006)}]{scharmer+spruit2006}
{Scharmer}, G.~B., \& {Spruit}, H.~C. 2006, \aap, 460, 605

\bibitem[{{Schlichenmaier} {et~al.}(2004){Schlichenmaier}, {Bellot Rubio}, \&
  {Tritschler}}]{schlichenmaier+bellot+tritschler2004}
{Schlichenmaier}, R., {Bellot Rubio}, L.~R., \& {Tritschler}, A. 2004, \aap,
  415, 731

\bibitem[{{Schlichenmaier} {et~al.}(2005){Schlichenmaier}, {Bellot Rubio}, \&
  {Tritschler}}]{schlichenmaier+bellot+tritschler2005}
---. 2005, Astronomische Nachrichten, 326, 301

\bibitem[{{Schlichenmaier} {et~al.}(1999){Schlichenmaier}, {Bruls}, \&
  {Sch{\"u}ssler}}]{schlichenmaier+bruls+schuessler1999}
{Schlichenmaier}, R., {Bruls}, J. H. M.~J., \& {Sch{\"u}ssler}, M. 1999, \aap,
  349, 961

\bibitem[{{Schlichenmaier} \& {Schmidt}(1999)}]{schlichenmaier+schmidt1999}
{Schlichenmaier}, R., \& {Schmidt}, W. 1999, \aap, 349, L37

\bibitem[{{Schlichenmaier} \& {Solanki}(2003)}]{schlichenmaier+solanki2003}
{Schlichenmaier}, R., \& {Solanki}, S.~K. 2003, \aap, 411, 257

\bibitem[{{Schmidt} \& {Balthasar}(1994)}]{schmidt+balthasar1994}
{Schmidt}, W., \& {Balthasar}, H. 1994, \aap, 283, 241

\bibitem[{{Sch{\"u}ssler} \& {V{\"o}gler}(2006)}]{schuessler+voegler2006}
{Sch{\"u}ssler}, M., \& {V{\"o}gler}, A. 2006, \apjl, 641, L73

\bibitem[{{Sigwarth} {et~al.}(1998){Sigwarth}, {Schmidt}, \&
  {Schuessler}}]{sigwarth+schuessler1998}
{Sigwarth}, M., {Schmidt}, W., \& {Schuessler}, M. 1998, \aap, 339, L53

\bibitem[{{Sobotka} \& {Hanslmeier}(2005)}]{sobotka+hanslmeier2005}
{Sobotka}, M., \& {Hanslmeier}, A. 2005, \aap, 442, 323

\bibitem[{{Sobotka} \& {Jurcak}(2008)}]{sobotka+jurcak2008}
{Sobotka}, M., \& {Jurcak}, J. 2008, 12th European Solar Physics Meeting,
  Freiburg, Germany, held September, 8-12, 2008.~Online at
  http://espm.kis.uni-freiburg.de/, p.2.23, 12, 2

\bibitem[{{Sobotka} {et~al.}(2008){Sobotka}, {Puschmann}, \&
  {Hamedivafa}}]{sobotka+puschmann+hamedivafa2008}
{Sobotka}, M., {Puschmann}, K.~G., \& {Hamedivafa}, H. 2008, Central European
  Astrophysical Bulletin, 32, 125

\bibitem[{{Socas-Navarro} {et~al.}(2004){Socas-Navarro}, {Pillet}, {Sobotka},
  \& {V{\'a}zquez}}]{socas+etal2004}
{Socas-Navarro}, H., {Pillet}, V.~M., {Sobotka}, M., \& {V{\'a}zquez}, M. 2004,
  \apj, 614, 448

\bibitem[{{Solanki} \& {Montavon}(1993)}]{solanki+montavon1993}
{Solanki}, S.~K., \& {Montavon}, C. A.~P. 1993, \aap, 275, 283

\bibitem[{{Solanki} {et~al.}(1994){Solanki}, {Montavon}, \&
  {Livingston}}]{solanki+montavon+livingston1994}
{Solanki}, S.~K., {Montavon}, C. A.~P., \& {Livingston}, W. 1994, \aap, 283,
  221

\bibitem[{{Solanki} \& {Schmidt}(1993)}]{solanki+schmidt1993}
{Solanki}, S.~K., \& {Schmidt}, H.~U. 1993, \aap, 267, 287

\bibitem[{{Spruit} \& {Scharmer}(2006)}]{spruit+scharmer2006}
{Spruit}, H.~C., \& {Scharmer}, G.~B. 2006, \aap, 447, 343

\bibitem[{{S{\"u}tterlin} {et~al.}(2004){S{\"u}tterlin}, {Bellot Rubio}, \&
  {Schlichenmaier}}]{suetterlin+bellot+schlichenmaier2004}
{S{\"u}tterlin}, P., {Bellot Rubio}, L.~R., \& {Schlichenmaier}, R. 2004, \aap,
  424, 1049

\bibitem[{{Title} {et~al.}(1993){Title}, {Frank}, {Shine}, {Tarbell}, {Topka},
  {Scharmer}, \& {Schmidt}}]{title+etal1993}
{Title}, A.~M., {Frank}, Z.~A., {Shine}, R.~A., {Tarbell}, T.~D., {Topka},
  K.~P., {Scharmer}, G., \& {Schmidt}, W. 1993, \apj, 403, 780

\bibitem[{{Tritschler} {et~al.}(2007){Tritschler}, {M{\"u}ller},
  {Schlichenmaier}, \& {Hagenaar}}]{tritschler+etal2007}
{Tritschler}, A., {M{\"u}ller}, D.~A.~N., {Schlichenmaier}, R., \& {Hagenaar},
  H.~J. 2007, \apjl, 671, L85

\bibitem[{{Tritschler} {et~al.}(2004){Tritschler}, {Schlichenmaier}, {Bellot
  Rubio}, {the KAOS Team}, {Berkefeld}, \& {Schelenz}}]{tritschler+etal2004}
{Tritschler}, A., {Schlichenmaier}, R., {Bellot Rubio}, L.~R., {the KAOS Team},
  {Berkefeld}, T., \& {Schelenz}, T. 2004, \aap, 415, 717

\bibitem[{{Westendorp Plaza} {et~al.}(2001){Westendorp Plaza}, {del Toro
  Iniesta}, {Ruiz Cobo}, \& {Mart{\'\i}nez Pillet}}]{westendorp+etal2001a}
{Westendorp Plaza}, C., {del Toro Iniesta}, J.~C., {Ruiz Cobo}, B., \&
  {Mart{\'\i}nez Pillet}, V. 2001, \apj, 547, 1148

\bibitem[{{Westendorp Plaza} {et~al.}(1997){Westendorp Plaza}, {del Toro
  Iniesta}, {Ruiz Cobo}, {Mart{\'\i}nez Pillet}, {Lites}, \&
  {Skumanich}}]{westendorp+etal1997}
{Westendorp Plaza}, C., {del Toro Iniesta}, J.~C., {Ruiz Cobo}, B.,
  {Mart{\'\i}nez Pillet}, V., {Lites}, B.~W., \& {Skumanich}, A. 1997, Nature,
  389, 47

\bibitem[{{Wiehr} \& {Degenhardt}(1993)}]{wiehr+degenhardt1993}
{Wiehr}, E., \& {Degenhardt}, D. 1993, \aap, 278, 584

\bibitem[{{Zakharov} {et~al.}(2008){Zakharov}, {Hirzberger}, {Riethm{\"u}ller},
  {Solanki}, \& {Kobel}}]{zakharov+etal2008}
{Zakharov}, V., {Hirzberger}, J., {Riethm{\"u}ller}, T.~L., {Solanki}, S.~K.,
  \& {Kobel}, P. 2008, \aap, 488, L17

\end{thebibliography}



\end{document}